\documentclass[final,twocolumn,pre,amsmath,amssymb,amsfonts,showpacs,superscriptaddress,titlepage,nobalancelastpage,raggedbottom,floatfix]{revtex4}

\usepackage[english]{babel}
\usepackage{ifpdf}
\ifpdf
\usepackage[pdftex]{graphicx}
\usepackage[protrusion=true,expansion=true]{microtype}
\usepackage[pdftex,colorlinks,linkcolor=black,citecolor=black,urlcolor=black,hyperfootnotes=false]{hyperref}
\else
\usepackage[dvips]{graphicx,color}
\usepackage{microtype}
\usepackage[dvips,colorlinks,linkcolor=black,citecolor=black,urlcolor=black,hyperfootnotes=false]{hyperref}
\fi

\newcommand{\integral}[4]{
\int_{#3}^{#4} #1 \,\mathrm{d}#2}

\newcommand{\e}[1]{ \mathrm{e}^{#1} }
\renewcommand{\exp}[1]{ \mathrm{exp}\left(#1\right) }

\newcommand{\remove}[1]{}

\begin{document}
\selectlanguage{english}
\title{Microtubule length distributions in the presence of protein-induced severing}
\author{Simon H. Tindemans}
\email{tindemans@amolf.nl} \affiliation{FOM Institute AMOLF,
Science Park 104, 1098 XG, Amsterdam, The Netherlands}
\affiliation{Max Planck Institute for the Physics of Complex
Systems, N\"othnitzer Strasse 38, 01187 Dresden, Germany}
\author{Bela M. Mulder}
\email{mulder@amolf.nl} \affiliation{FOM Institute AMOLF, Science
Park 104, 1098 XG, Amsterdam, The Netherlands}
\date{\today}
\begin{abstract}
Microtubules are highly regulated dynamic elements of the
cytoskeleton of eukaryotic cells. One of the regulation mechanisms
observed in living cells is the severing by the proteins katanin
and spastin. We introduce a model for the dynamics of microtubules
in the presence of randomly occurring severing events. Under the
biologically motivated assumption that the newly created plus end
undergoes a catastrophe, we investigate the steady state length
distribution. We show that the presence of severing does not
affect the number of microtubules, regardless of the distribution
of severing events. In the special case in which the microtubules
cannot recover from the depolymerizing state (no rescue events) we
derive an analytical expression for the length distribution. In
the general case we transform the problem into a single ODE that
is solved numerically.
\end{abstract}


\maketitle

\section{Introduction}

Microtubules are filamentous protein aggregates that appear in all
eukaryotic cells. They have an inherent polarity that results in
different dynamics at their two ends. The so-called plus end is
highly dynamic, alternating between prolonged periods of
polymerization (growth) and depolymerization (shrinkage)
\cite{Mitchison1984}. On the other end of the filament, the minus
end often remains connected to the locus of nucleation
\cite{Burk2007} or is found to exhibit relatively steady
depolymerization \cite{Shaw+03}. The combination of slow
depolymerization at the minus end and prolonged growth at the plus
end leads to a phenomenon known as treadmilling, whereby the
individual tubulin dimers appear to move from the plus to the
minus end \cite{Rodionov1997}. As the stiffest of the cytoskeletal
filaments, microtubules are widely used in intracellular transport
and for structural support. Therefore, the dynamic properties of
the microtubules are tightly regulated by the cell, often through
the use of microtubule-associated proteins (MAPs) \cite{Alberts}.

In this work we investigate one particular type of MAP that causes
the severing of microtubules. It was noted by Vale \cite{Vale1991}
that otherwise stable microtubules could be severed in mitotic
extracts of Xenopus eggs. This activity was traced back to a
protein that is able to use ATP hydrolysis to sever microtubules.
The protein was identified only later and given the name katanin
after the katana, the Japanese Samurai sword \cite{McNally1993}.
Katanin is a heterodimer, consisting of the p60 and p80 subunits:
the p80 subunit is thought to be responsible for the targeting of
the protein, whereas p60 is involved in the actual severing as
part of a hexameric ring \cite{Hartman1999}. The hexameric form of
katanin appears to remove individual dimers from the microtubule
lattice, thereby compromising the structural integrity of the
microtubule. It is not currently clear whether katanin acts
uniformly along the microtubule, or whether it is attracted by
pre-existing lattice defects \cite{Davis02}.

Katanin homologs have since been discovered across the animal and
plant kingdoms \cite{Burk2007}. Another severing protein by the
name of spastin has also been identified. Like katanin, it also
assembles in a hexameric ring, suggesting a severing mechanism
similar to that of katanin \cite{Roll-Mecak2008}. Both katanin and
spastin are capable of severing microtubules at seemingly random
locations, and are used by the cell for the regulation of the
cytoskeleton \cite{rollmecak2006}, for example in the mitotic and
meiotic spindles \cite{McNally2006} or the formation of the
transverse cortical array in plant cells \cite{Burk2001,
Wightman07}. Generally, the activity of severing proteins leads to
a decrease in the average microtubule length, but an increase in
their number \cite{rollmecak2006}. Surprisingly, this increase in
number can sometimes more than offset the loss of microtubule
length due to the average length decrease \cite{Srayko2006}.

In this work, we investigate theoretically how the occurrence of
microtubule severing at random positions affects the length
distribution of microtubules. Previous studies have assessed the
effect of severing on actin filaments
\cite{Edelstein1998,Roland2008}. However, whereas actin has a
single growth mode that is described well by constant
polymerization and depolymerization rates, microtubules show
richer dynamics because the plus end switches between growing
(polymerizing) and shrinking (depolymerizing) states.

\section{Model}

We base our model on the basic dynamic instability model that was
introduced by Dogterom and Leibler \cite{Marileen93}. In this
model, microtubules exist in either the growing or the shrinking
state, and switch between these states with a `catastrophe' rate
$r_c$ (growth $\to$ shrinkage) and a `rescue' rate $r_r$
(shrinkage $\to$ growth). In the growing state, the microtubule
extends with an effective speed $v^+$, and in the shrinking state
it retreats with an effective speed $v^-$. New microtubules are
nucleated with a steady nucleation rate $r_n$. This set of
constraints gives rise to a steady state distribution of
microtubule lengths, provided that $v^+/r_c < v^-/r_r$.

We model microtubule severing by a constant severing rate per unit
length. This is a valid approximation if we assume that
microtubule severing process occurs on a time scale that is much
shorter than the time in which a microtubule grows significantly.
By taking a constant severing rate, we also implicitly assume that
severing is limited by the availability of microtubules, i.e. the
severing protein is available in abundance. However, because we
focus on steady-state results, where the total amount of
microtubules is constant, the possible invalidity of this
assumption will not qualitatively affect the results.

Following the approach in \cite{Marileen93}, we construct a set
of evolution equations for the length distributions of growing and
shrinking microtubules. Denoting the growing and shrinking
microtubule length distributions by $m^+(l,t)$ and $m^-(l,t)$,
respectively, the equations can be written as
\begin{subequations}\label{eq:sevEqnAll}
\begin{align}
\frac{\partial }{\partial t}m^+(l,t)=& -v^+ \frac{\partial }{\partial l}m^+(l,t) -r_c m^+(l,t) + r_r m^-(l,t) \nonumber\\ &+ \Phi^+_{\text{\textit{severing}}} \label{eq:sevEqnAll-1}\\
\frac{\partial }{\partial t}m^-(l,t)=& v^-
\frac{\partial}{\partial l} m^-(l,t) + r_c m^+(l,t) - r_r m^-(l,t)
\nonumber\\ &+
\Phi^-_{\text{\textit{severing}}}\label{eq:sevEqnAll-2}\mathrm{,}
\end{align}
\end{subequations}
where the derivatives with respect to $l$ reflect the translation
of the distributions due to growth and shrinkage of microtubules.
The severing contribution $\Phi_{\text{\textit{severing}}}$ will
be constructed below. These equations are supplemented by the
boundary condition
\begin{equation}
m^+(0,t) = \frac{r_n}{v^+},
\end{equation}
specifying the nucleation of new microtubules with rate $r_n$, and
by the physically motivated constraint that both distributions
tend to zero for large lengths (there are no infinitely long
microtubules).

\begin{table}[t]
\begin{tabular}
[c]{ll}\hline Model parameters&\\\hline
$v^{+}$ & growth speed\\
$v^{-}$ & shrinkage speed\\
$r_{c}$ & catastrophe rate\\
$r_{r}$ & rescue rate\\
$r_{n}$ & nucleation rate\\
$r_{s}$ & severing rate\\
\hline Dimensionless parameters&\\
\hline
$v = v^+/v^-$ & speed ratio\\
$r = r_r/r_c$ & rescue rate\\
$s = r_s v^+/r_c^2 $& severing rate\\
\hline
\end{tabular}
\caption{Overview of the parameters in natural
dimensions and the dimensionless parameters}\label{tab:severing-parameters}
\end{table}

In the following we derive the form of the severing terms in the
set of equations \eqref{eq:sevEqnAll}. The inclusion of severing
events leads to two types of contributions to the evolution of the
length distributions: the \emph{disappearance} flux
$\phi_{\textrm{out}}(l,t)$ of microtubules of a certain length $l$
and the \emph{appearance} fluxes $\phi_{\textrm{in}}(l',t)$ and
$\phi_{\textrm{in}}(l'',t)$ of two new microtubules with a total
length $l=l'+l''$. In addition, the newly created microtubules can
be created from microtubules that were initially growing (+) or
shrinking (-). Symbolically, we write
\begin{align}
\Phi^+_{\text{severing}} &= - \phi^+_{\textrm{out}}(l,t) +
\phi^+_{\textrm{in},+}(l,t) + \phi^+_{\textrm{in},-}(l,t)\\
\Phi^-_{\text{severing}} &= - \phi^-_{\textrm{out}}(l,t) +
\phi^-_{\textrm{in},+}(l,t) + \phi^-_{\textrm{in},-}(l,t),
\end{align}
where $\phi^{\sigma}_{\textrm{out}}(l,t)$ stands for the removal of microtubules in state $\sigma \in \{-,+\}$, and $\phi^{\sigma}_{\textrm{in},\tau}(l,t)$ represents the appearance of new microtubules of length $l$ in state $\sigma$ as a result of the severing of microtubules in state $\tau$. The contributions will be discussed individually below.

The process of severing is controlled by the severing rate $r_s$
that is given as a rate per unit of length. Thus we find that the
fluxes of disappearing microtubules $\phi^+_{\textrm{out}}(l,t)$
(growing) and $\phi^-_{\textrm{out}}(l,t)$ (shrinking) are given
by
\begin{align}
\phi^+_{\textrm{out}}(l,t) &= r_s l m^+(l,t)\\
\phi^-_{\textrm{out}}(l,t) &= r_s l m^-(l,t).
\end{align}

In order to specify the influx terms
$\phi^{\pm}_{\textrm{in},\pm}$ it is necessary to specify the
process of severing in more detail. We will assume that the action
of the severing protein is local, thus having no effect on the
remote plus and minus ends of the microtubule it severs. This
implies that the plus end fragment of a severed microtubule
remains in the same state. However, we must make an explicit
assumption regarding the state of the newly created plus end. In
line with biological observations \cite{Quarmby2000}, we assume
that this plus end always starts out in the shrinking state. A
severing event thus shortens an existing microtubule without
affecting its growth state and creates an additional microtubule
that is in the shrinking state. If necessary, the model can easily
be extended to handle (a fraction of) severing-created plus ends
that start out in the growing state.

Let us define the uniform probability density of selecting a
severing location $l$ on a microtubule of length $l'$ as
$p(l|l')$, with the value $1/l'$ for $l\in[0,l']$, and $0$
otherwise. We then derive the influx
$\phi^{+}_{\textrm{in},+}(l,t)$ of growing microtubules of length
$l$ that results from the severing of growing microtubules.
\begin{align}
\phi^{+}_{\textrm{in},+}(l,t) &= \integral{\phi^+_{\textrm{out}}(l',t) p(l'-l|l')}{l'}{0}{\infty}\nonumber\\
&= r_s\integral{l' m^+(l')\theta(l'-l)\frac{1}{l'}}{l'}{0}{\infty}\nonumber\\
&= r_s \integral{m^+(l')}{l'}{l}{\infty},
\end{align}
where $\theta(x)$ is the Heaviside step function. In a similar
way, and taking account the fact that the minus end fragment is always in a shrinking state, we derive
\begin{align}
\phi^{-}_{\textrm{in},+}(l,t)&= \integral{\phi^+_{\textrm{out}}(l',t) p(l|l')}{l'}{0}{\infty}\nonumber\\
&= r_s \integral{ m^+(l')}{l'}{l}{\infty}\\
\phi^{-}_{\textrm{in},-}(l,t)&= \integral{\phi^-_{\textrm{out}}(l',t) \left[ p(l|l') + p(l'-l|l') \right]}{l'}{0}{\infty}\nonumber\\
&= 2 r_s \integral{
m^-(l')}{l'}{l}{\infty}.
\end{align}
And, because the severing of shrinking microtubules cannot produce growing microtubules, we have $\phi^{+}_{\textrm{in},-}(l,t)= 0$.

Note that we do not need to keep track of the correlations between
the lengths of the \emph{individual} microtubules that are created
from a single cutting event, because we are looking only at
ensemble-averaged length distributions. The contributions derived
above are similar in form to those introduced by Edelstein-Keshet
and Ermentrout \cite{Edelstein1998} as the continuous limit of a
discrete monomer addition and severing model for actin filaments.
However, we use distinct asymmetric terms for growing and
shrinking microtubules.

The model as described does not explicitly take
into account the often observed depolymerization of microtubules at their minus
ends that leads to treadmilling. However, this case is easily
addressed through a renormalization of the growth and shrinkage
speeds. Denoting the shrinkage speed at the minus end by $v^{tm}$,
this is achieved by the substitutions $v^+ \to v^+ - v^{tm}$ and
$v^- \to v^- +v^{tm}$, leaving the results qualitatively unchanged by the incorporation of treadmilling.

\subsection{Steady state equations}

Inserting the severing terms into equations \eqref{eq:sevEqnAll} we obtain
\begin{subequations}\label{eq:sevEqnFull}
\begin{align}
\frac{\partial}{\partial t} m^+(l,t)=&-r_c m^+(l,t) + r_r m^-(l,t) -v^+ \frac{\partial}{\partial l} m^+(l,t)\nonumber\\& - r_s l m^+(l,t)+
r_s \integral{m^+(l',t)}{l'}{l}{\infty} \\
\frac{\partial}{\partial t} m^-(l,t)=&+r_c m^+(l,t) - r_r m^-(l,t) +v^- \frac{\partial}{\partial l} m^-(l,t) \nonumber\\&- r_s l m^-(l)+
r_s \integral{[m^+(l') + 2 m^-(l')]}{l'}{l}{\infty} \mathrm{,}
\end{align}
\end{subequations}
We note that in the special case when $r_c=r_r=0$, the equation for the length distribution of the growing microtubules is equivalent to the model of microtubule GTP-cap dynamics by Flyvbjerg \emph{et al.} \cite{Flyvbjerg04} with a vanishing diffusion constant.

The number of parameters in these equations can be reduced by scaling the parameters and variables using natural units for length, time and microtubule number. As a unit of time, we take the mean time to catastrophe ($1/r_c$); the unit of length is the microtubule growth in that time ($v^+/r_c$) and unit of microtubule number is the number of microtubules nucleated in this time ($r_n/r_c$). In terms of the dimensionless length unit $x= \frac{r_c}{v^+} l$, we define \begin{align}
f^+(x) &\equiv \frac{v^+}{r_n}m^+(l(x),t)\\
f^-(x) &\equiv \frac{v^+}{r_n}m^-(l(x),t)\\
v &\equiv v^+/v^-\\
r &\equiv r_r/r_c\\
s &\equiv r_s v^+/r_c^2 \mathrm{.}
\end{align}
Note that the definition of $v$ is inverted with respect to its
natural conversion into dimensionless parameters, but this choice
simplifies the notation in what follows. In the steady state, the
equations \eqref{eq:sevEqnFull} in dimensionless form become
\begin{subequations}\label{eq:sever-SSall}
\begin{align}
\frac{\mathrm{d}}{\mathrm{d}x} f^+(x)=&- f^+(x) + r f^-(x)  - s x f^+(x)\nonumber\\&+
s \integral{f^+(x')}{x'}{x}{\infty} \label{eq:sseqn1}\\
\frac{1}{v}\frac{\mathrm{d}}{\mathrm{d}x} f^-(x)=&- f^+(x) + r f^-(x)  + s x f^-(x) \nonumber\\&- s
\integral{ [f^+(x') + 2 f^-(x')]}{x'}{x}{\infty} \mathrm{,}
\label{eq:sseqn2}
\end{align}
\end{subequations}
with the boundary conditions
\begin{equation} \label{eq:ssboundary}
f^+(0) = 1 \quad ; \quad
\lim_{x \to \infty}f^+(x) = 0 \quad ; \quad \lim_{x \to
\infty}f^-(x) = 0.
\end{equation}
In the absence of severing ($s=0$) these equations are solved by
the exponential functions $f^+(x) = \mathrm{e}^{-(1-r v)x}$ and
$f^-(x) = v \mathrm{e}^{-(1-r v)x}$ \cite{Marileen93}. Note that
these solutions are only valid for $r v < 1$, when the average
microtubule length is bounded.

We proceed to analyze the steady state equations
\eqref{eq:sever-SSall} in a number of steps. First, we determine
global properties of the length distributions. Subsequently, we
derive an explicit expression for the microtubule length
distributions in the special case in which rescues are absent ($r
= 0$). We conclude with a numerical method using which the
distribution can be calculated for arbitrary parameter values.

\section{Results}

\subsection{Number of microtubules}

Multiplying equations \eqref{eq:sever-SSall} by $x$, integrating
from $0$ to $\infty$ and subtracting the results yields
\begin{equation} \label{eq:plusminrelation}
v \integral{ f^{+}(x)}{x}{0}{\infty} =
\integral{f^{-}(x)}{x}{0}{\infty}.
\end{equation}
Also, integrating over \eqref{eq:sseqn1} yields
\begin{equation}
-f^+(0) = - \integral{ \left[ f^+(x) - r f^-(x)\right] }{x}{0}{\infty},
\end{equation}
which can be combined with \eqref{eq:ssboundary} and \eqref{eq:plusminrelation} to give the following expression for the total number of microtubules.
\begin{equation} \label{eq:totalnumber}
\integral{(f^{+}(x) + f^{-}(x))}{x}{0}{\infty} = \frac{1+v}{1-r v}
\end{equation}
Surprisingly, the total number of microtubules in the system does
\emph{not} depend on the rate of microtubule severing. Specifically, we note that severing can not prevent the diverging microtubule count for $r v \to 1$.

\subsection{In the absence of rescue events}\label{sec:norescue}
\begin{figure}[ht]
\begin{center}
\includegraphics{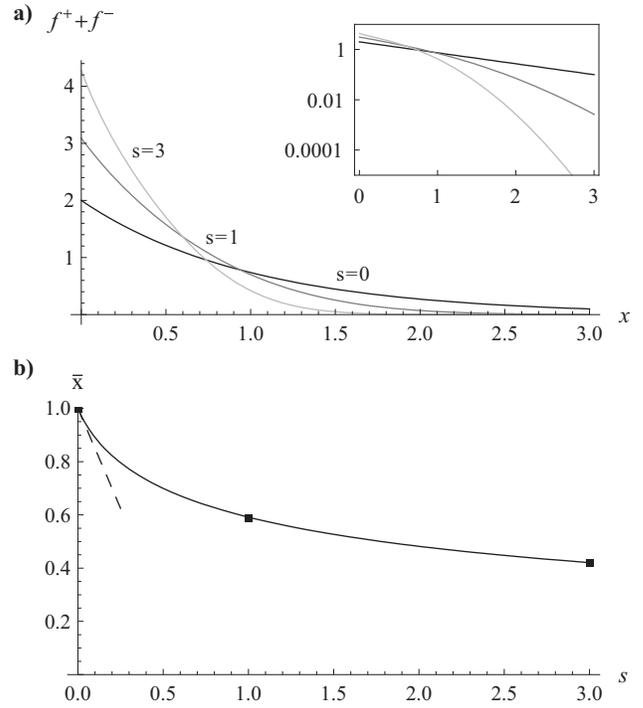}
\caption{Results in the absence of rescues ($v=1/2$, $r=0$). The
choice $v=1/2$ is an approximate value based on the \emph{in-vivo}
measurements in interphase plant cells reported by Vos \emph{et
al.} \cite{vos04}. \textbf{a)} Length distributions for three
different values of the severing rate: $s=0$ (black), $s=1$ (dark
gray) and $s=3$ (light gray). The inset shows the same on a
logarithmic scale. \textbf{b)} Average length as a function of the
severing parameter $s$. The squares indicate the parameter values
of the distributions in (a). The dashed line is the small-$s$
approximation \eqref{eq:smallk-x}. Length distributions have been
calculated from \eqref{eq:FplusR0} and \eqref{eq:FminR0}; the
average length from \eqref{eq:avLengthR0}.}\label{fig:severingR0}
\end{center}
\end{figure}

In the special case $r=0$ (no rescue events), equation
\eqref{eq:sseqn1} for $f^+(x)$ decouples from \eqref{eq:sseqn2}
and can be solved analytically. This solution can then be used to
obtain an expression for $f^-(x)$ as well. We introduce the
functions $F^{\pm}(x)=-\integral{f^{\pm}(x')}{x'}{x}{\infty}$,
allowing us to rewrite equations \eqref{eq:sever-SSall} as
\begin{align}
\frac{\mathrm{d}}{\mathrm{d}x} (\frac{\mathrm{d}}{\mathrm{d}x} F^+(x) +& (1+s x)F^+(x))=0 \label{eq:sever-primeq1}\\
\frac{1}{v}\frac{\mathrm{d}^2}{\mathrm{d}x^2} F^-(x) =& -\frac{\mathrm{d}}{\mathrm{d}x} F^+(x)+s x \frac{\mathrm{d}}{\mathrm{d}x}
F^-(x)\nonumber\\&+s F^+(x) +2 s F^-(x)\label{eq:sever-primeq2}
\end{align}
From equations \eqref{eq:plusminrelation} and \eqref{eq:totalnumber} we find that $F^+(0)=-1$ (because $r=0$), and using the boundary condition $\partial_x F^+(0) = f^+(0)=1$ we solve \eqref{eq:sever-primeq1} to obtain
\begin{equation}
F^+(x)= - \exp{-x-\frac{1}{2}s x^2}.\label{eq:FplusR0}
\end{equation}
Inserting this solution into equation \eqref{eq:sever-primeq2}, it
can be solved using the boundary conditions $F^-(0)=-v$ (from equation
\eqref{eq:totalnumber}) and $\lim_{x\to \infty}F^-(0)=0$ to give
\begin{align}
F^-(x)=& -v \, \exp{-x-\frac{1}{2}s v^2} \times\nonumber\\
&\biggr[ 1-\sqrt{\frac{\pi}{2}}\sqrt{s(1+v)}x \,
\exp{\frac{(1+s(1+v)x)^2}{2s(1+v)}}\times \nonumber\\
& \mathrm{erfc}\left(\frac{1+s(1+v)x}{\sqrt{2s(1+v)}}\right)
 \biggr],\label{eq:FminR0}
\end{align}
where $\mathrm{erfc}(z)$ is the complementary error function
\begin{equation}
\mathrm{erfc}(y) = \frac{2}{\sqrt{\pi}}
\integral{\e{-t^2}}{t}{y}{\infty}.
\end{equation}
The expressions for $f^+(x)$ and $f^-(x)$ follow by
differentiation of \eqref{eq:FminR0}. The resulting distribution
for $v=1/2$ and various values of $s$ is shown in figure
\ref{fig:severingR0}.

Looking at the derivative of the combined distribution
$f^+(x)+f^-(x)$, we find that for $x=0$
\begin{equation}
\frac{\mathrm{d}}{\mathrm{d}x} (f^+(x)+f^-(x))|_{x=0} = -(1+v)(1-s(1-2v)).
\end{equation}
The perhaps surprising implication is that if $v<1/2$,
sufficiently high severing rates can lead to a positive slope of
the distribution at $x=0$. In other words, in that case the length
distribution no longer decreases monotonically with length.

Finally, we compute the average microtubule length as
\begin{align}
\bar{x}&=\frac{\integral{x(f^+(x)+f^-(x))}{x}{0}{\infty}}{\integral{(f^+(x)+f^-(x))}{x}{0}{\infty}}\\
&= -\frac{1}{1+v}\integral{\left( F^+(x)+F^-(x)\right)}{x}{0}{\infty}\\
&=\frac{2}{z}\e{1/z^2}\integral{\e{-t^2}}{t}{1/z}{\infty},
\label{eq:avLengthR0}
\end{align}
with $z = s(1+v)$. This confirms that the average length decreases
with increasing severing activity, a fact that is reflected in
figure \ref{fig:severingR0}.

\subsection{Numerical evaluation of the length distribution}

For the more general case $r > 0$ we use a numerical method to solve equations \eqref{eq:sever-SSall}. In order to do so, we reparameterize the problem. Inspired by the results \eqref{eq:plusminrelation} and \eqref{eq:totalnumber} we
define
\begin{eqnarray}
p(x) &= & f^+(x)+f^-(x)\\
q(x) &=& v f^+(x) - f^-(x) \mathrm{.} \label{eq:qdefinition}
\end{eqnarray}
Hence, $p(x)$ is a (dimensionless) measure of the total
microtubule density. In terms of these variables, the steady state equations \eqref{eq:sever-SSall} can be written as
\begin{eqnarray}
\frac{\mathrm{d}}{\mathrm{d}x} q(x) &=& s v \left[ -x p(x)+2\int_x^{\infty} p(x')\mathrm{d}x'
\right] \label{eq:divqdef}\\
\frac{\mathrm{d}}{\mathrm{d}x} p(x) &=& -(1-r v)p(x)-(1+r)q(x)\nonumber\\
&&-s(1-v)x p(x)-s x q(x)+\nonumber\\
&& s \int_x^{\infty} \left[q(x')+(1-2v)p(x') \right] \mathrm{d}x'.
\label{eq:divpdef}
\end{eqnarray}
Equation \eqref{eq:divqdef} can be formally solved to give
\begin{align}
q(x) &= s v \int_x^{\infty}\left[ x' p(x') -2 \int_{x'}^{\infty}
p(x'')\mathrm{d}x'' \right] \mathrm{d}x' \\
&= s v \int_x^{\infty}\left[ (2x-x') p(x') \right] \mathrm{d}x'
\mathrm{.} \label{eq:qprelation}
\end{align}
Inserting this expression into equation \eqref{eq:divpdef} we
transform the problem into a single integral equation. We can
subsequently remove the integrals by differentiating three times
and obtain the linear fourth-order ODE
\begin{align}
p^{(4)}(x)=& (-1+r v-s(1-v)x) p^{(3)}(x)\nonumber\\&
+s(-4+v(5+x(1+r+s
x)))p^{(2)}(x) \nonumber\\
&+ 4 s v(1+r+2s x)p^{(1)}(x) + 12 v s^2 p(x)\label{eq:severODE},
\end{align}
where $p^{(n)}(x)$ stands for $(\mathrm{d}/\mathrm{d}x)^n p(x)$.
To derive the boundary conditions for this problem we use \eqref{eq:totalnumber},\eqref{eq:plusminrelation} and \eqref{eq:ssboundary} to derive
\begin{align}
\integral{p(x)}{x}{0}{\infty} &= \frac{1+v}{1-r v},\label{eq:sever-ODE-check1}\\
\integral{q(x)}{x}{0}{\infty} &=0,\\
q(0) &=-p(0) + 1+v.
\end{align}
By repeated application of these equalities and differentiation of
equations \eqref{eq:divqdef} and \eqref{eq:divpdef} we obtain the
boundary conditions
\begin{align}
p^{(1)}(0) =& (1+v)r p(0)-(1+v)(1+r) \nonumber\\&+ s(1+v)\left(\frac{1-2v}{1-r v}\right)\\
p^{(2)}(0) =& \left[ -r(1+v)(1-r v)+3 s v \right]p(0) +(1+v)\times\nonumber\\ &\left[(1+r)(1-r v)-s\left( \frac{3-r v+2 r v^2}{1-r v} \right) \right]\\
p^{(3)}(0) =& r(1+v)\left[ (1-r v)^2 + s r(-3+7v) \right]p(0) \nonumber\\ &- (1+v)(1+r)(1-r v)^2 \nonumber\\ &+
 s(1+v)\left[ 6+3r-4v-5r v+2r v^2 \right]\nonumber\\ & - s^2\left(
\frac{1+v}{1-r v} \right)(3-4v+8v^2).
\end{align}
We note that these boundary conditions are of the form
\begin{equation}
\left(\begin{array}{c} p(0)\\ p^{(1)}(0)\\p^{(2)}(0)\\p^{(3)}(0)
\end{array}\right) = \mathbf{P_1} p(0) + \mathbf{P_2}
\end{equation}
with a single undetermined parameter $p(0)$. Because equation
\eqref{eq:severODE} is a homogeneous linear ODE, we can evaluate
it twice, using both $\mathbf{P_1}$ and $\mathbf{P_2}$ as boundary
conditions. We generally find that both solutions diverge with
opposing signs. The value of $p(0)$ can therefore be determined from the
constraint $\lim_{x\to \infty}p(x)=0$.

\begin{figure}[ht]
\begin{center}
\includegraphics{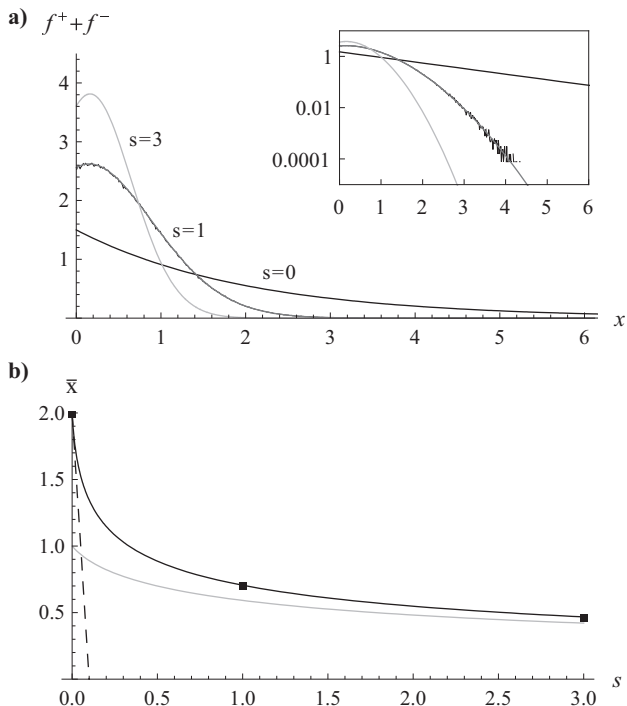}
\caption{Numerically
computed length distributions and comparison with
simulation data. Based on the \emph{in-vivo}
measurements in interphase plant cells reported by Vos \emph{et al.} \cite{vos04}, approximate values of  $v=1/2$ and $r=1$ have been used. \textbf{a)} Length distributions for three
different values of the severing rate: $s=0$ (black), $s=1$ (dark
gray) and $s=3$ (light gray). The inset shows the same on a
logarithmic scale. Also shown for $s=1$ is data obtained by a stochastic simulation of microtubule severing using a Gillespie algorithm, matching the predicted distribution. Simulation parameters were [$v^+ = 0.1\,\mu$m
s$^{-1}$, $v^- = 0.2\,\mu$m s$^{-1}$, $r_c=0.01 \,$s$^{-1}$,
$r_r=0.01 \,$s$^{-1}$, $r_n=10 \,$s$^{-1}$, $r_s=0.001
\,\mu$m$^{-1}$s$^{-1}$]. Length data was distributed into 500 bins
and sampled 1000 times at 50 second intervals after an initial
equilibration period of 50,000 seconds. \textbf{b)} Average length
as a function of the severing parameter $s$. The squares indicate
the parameter values of the distributions in (a). The dashed line
is the small-$s$ approximation \eqref{eq:smallk-x} The light gray
curve, which converges for large $s$, is the result for $r=0$
(figure \ref{fig:severingR0}). Length distributions
have been calculated from \eqref{eq:severODE}.}\label{fig:severingR1}
\end{center}
\end{figure}

To this end, both solutions should be evaluated over a range that
is as large as possible, whilst maintaining a very high numerical
accuracy, because the final result is obtained by subtracting
two diverging functions. In our numerical calculations (using Wolfram Mathematica 6.0), we have evaluated the differential equation with sufficient precision to achieve an accuracy of 13 significant digits and limited the integration range to 10 decay lengths $(x=10(1-r v)^{-1})$ \emph{or} the point at which the first solution exceeded the value $10^8$, depending on which occurred first.

\begin{figure}[ht]
\begin{center}
\includegraphics{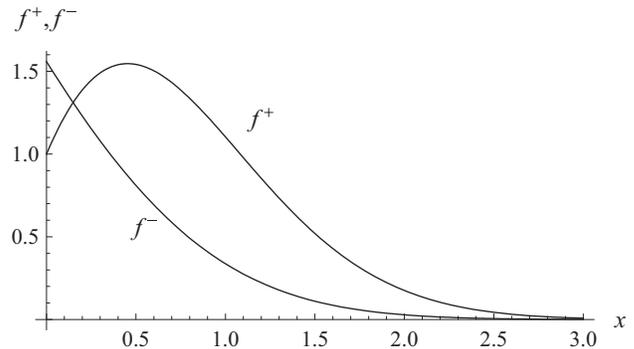}
\caption{Length distributions for
growing and shrinking microtubules separately ($v=1/2$, $r=1$,
$s=1$). The total distribution function $p(x)=f^+(x)+f^-(x)$ has
been computed numerically using \eqref{eq:severODE}; $f^+(x)$ and
$f^-(x)$ have been computed using \eqref{eq:qprelation} and
\eqref{eq:qdefinition}.}\label{fig:fplusminus}
\end{center}
\end{figure}

Figure \ref{fig:severingR1} shows the numerically computed
distributions for $v=1/2$, $r=1$ and various values of $s$. It is
interesting to note that the total distribution is no longer
monotonically decreasing for $s=1$ and $s=3$. Figure
\ref{fig:fplusminus} shows that this is solely due to the
contribution from the growing microtubules. We also note that the
average length decreases rapidly for relatively small severing
rate. As the severing rate increases, the average length converges
to that of the system without rescue events ($r=0$). In other
words, if severing events occur very frequently, rescue events are
no longer significant, presumably because the microtubules become
so short that they disappear before they can be rescued. This
statement is summarized by the condition $1/r_r \gg \bar{l}/v^-$, where $\bar{l}$ is the average microtubule length, or in dimensionless units $r v \bar{x} \ll 1$.

\section{Discussion}\label{sec:sev-discussion}

We have constructed a model that describes the dynamic instability
of microtubules in combination with the severing of microtubules.
This model takes the form of two coupled integro-differential
equations that are a function of three parameters: $v$, the ratio
of polymerization and depolymerization speeds, $r$, the ratio of
the rescue and catastrophe rates, and the dimensionless severing
rate $s$. We have analyzed the steady state solutions and their
properties: notably the number of microtubules and their average
length. For the special cases of no rescue events ($r=0$) and
small severing rates (see appendix \ref{sec:smallsev}), we have
presented analytical solutions. The general case has been
addressed by transforming the coupled integro-differential
equations into a single fourth order differential equation that is
solved numerically. The resulting microtubule length distributions
show a number of interesting properties.

\subsubsection{Shorter, more compact length distributions}

As expected, an increase in the severing rate always leads to a
decrease in the average length of the microtubules. In addition,
because the rate of severing is proportional to the microtubule
length, the number of very long microtubules is strongly reduced,
leading to a distribution that is more compact (see appendix
\ref{sec:smallsev} for an explicit expression). Furthermore, in
contrast to the dynamic instability model without severing, the
length distributions are no longer guaranteed to decrease
monotonically with increasing length. For some parameters, a
`bump' is observed in the length distribution, caused by the
continuous creation of short (but not vanishingly small)
microtubules through the severing process.

\subsubsection{Conservation of microtubule number}

We have also demonstrated that the total number of microtubules
does not depend on the severing rate, even though the average
length of each microtubule decreases. To understand this
counterintuitive result, we consider a system without severing, in
which the following steady-state relation holds: [population size]
$=$ [nucleation rate] $\times $ [average lifetime]. Suppose that a
single microtubule is severed (consistent with an infinitesimally
small severing rate). The expected lifetime of the created
segments is obviously shorter than that of the original segment.
However, the fact that the total number of microtubules is
unaffected by the presence of severing implies that this decrease
in lifetime is compensated \emph{exactly} by the fact that every
severing event `nucleates' an extra microtubule. For this
cancellation to occur, the sum of the lifetimes of the two
fragments of a severed microtubule should equal the expected
lifetime of the microtubule if severing had not occurred.

\begin{figure}[ht]
\begin{center}
\includegraphics{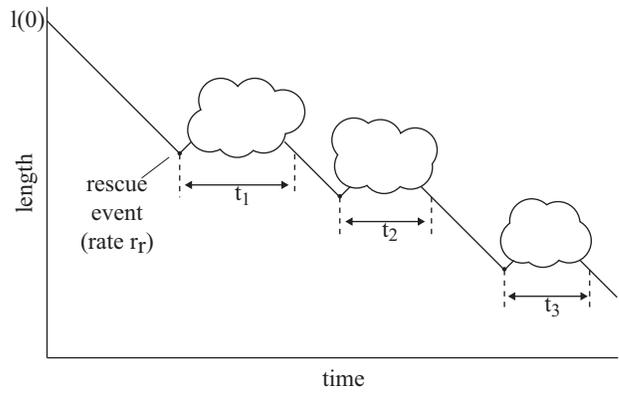}
\caption{Schematic depiction of the life history of a shrinking microtubule with an initial length $l(0)$. The return times $t_i$ have an expectation value $T^+$.}\label{fig:shrinkdynamics}
\end{center}
\end{figure}

That this is indeed the case can be shown using a simple argument,
based on the lifetime of a microtubule in the absence of severing.
In this case the dynamics of the microtubule tip are independent
of its length. This implies that the average time it takes for a
growing microtubule of length $l$ to later return (in a shrinking
state) to the \emph{same} length is independent of the value of
$l$, and therefore equal to the average microtubule lifetime $T^+$
(obtained by starting from $l(t_0)=0$). The remaining contribution
to the lifetime is given by the time it takes for the shrinking
microtubule of length $l$ to disappear. In the absence of rescue
events, this would take a time $l/v^-$. However, each rescue event
switches the microtubule back to the growing state, extending the
microtubule's expected lifetime by $T^+$ before it returns to the
same position in the shrinking state. The expected number of such
rescue events is equal to $r_r l/v^-$. This argument is
illustrated in figure \ref{fig:shrinkdynamics}. Collecting the
terms described above, we obtain the following expression for the
expected microtubule lifetime $T(l,\sigma)$ of a microtubule of
length $l$ and growth state $\sigma \in \{-,+\}$:
\begin{equation}
T(l,\sigma) = T^+ \delta_{\sigma,+} + \frac{l}{v^-}\left( 1 + r_r
T^+ \right).
\end{equation}
The same relation, including an expression for $T^+$, has been
derived analytically by and Bicout \cite{Bicout97} and an
equivalent expression for microtubules that grow in discrete units
was produced by Rubin \cite{Rubin88}.

Each severing event conserves the total length $l$ of the severed
microtubule and the state $\sigma$ of the existing plus end, and
the newly created plus end immediately undergoes a catastrophe
($\sigma_{\mathrm{new}}=-$). In terms of life times we find that
$T(l,\sigma)=T(l',\sigma) + T(l-l',-)$, so that a single severing
event indeed preserves the total lifetime. By induction it follows
that \emph{any} number of severing events will leave the expected
number of microtubules in the system intact.

We stress that this argument holds regardless of the frequency and
location of severing events. This implies that any distribution of
severing events -- provided they lead to a catastrophe of the
trailing end -- conserves the total number of microtubules in
steady state. Specifically, this also applies to severing at
positions where microtubules of different orientations cross
\cite{Wightman07}, a process that may be relevant to the formation
of the cortical array in plant cells.

The conservation of microtubule number in the presence of severing
is in apparent contradiction with the experimentally reported
increase in microtubule numbers \cite{rollmecak2006}. However,
this can be readily explained by the fact that our results are based on the assumption of
constant parameter values. In a living cell, the decrease in
average microtubule length that is the result of severing will
lead to an increased availability of free tubulin dimers. In turn,
this is likely to increase the polymerization rate (growth speed)
and nucleation rate, which would indeed cause an increase in the number
of microtubules.

\begin{acknowledgments}
ST thanks Frank J\"ulicher and Benjamin Lindner for very
helpful discussions and is grateful for the hospitality that was
provided by the Max Planck Institute for the Physics of Complex
Systems. The authors thank Nils Becker for critical comments on the manuscript. ST was supported by a grant from the NWO ``Computational
Life Sciences'' program (contract CLS 635.100.003) and a visiting
stipend from the Max Planck Society. This work is part of the
research program of the ``Stichting voor Fundamenteel Onderzoek
der Materie (FOM)'', which is financially supported by the
``Nederlandse organisatie voor Wetenschappelijk Onderzoek (NWO)''.
\end{acknowledgments}

\appendix

\section{Small severing rates} \label{sec:smallsev}
In the absence of severing, equations \eqref{eq:sever-SSall} are
solved by $f^+(x) = \mathrm{e}^{-(1-r v)x}$ and $f^-(x) = v
\mathrm{e}^{-(1-r v)x}$. To investigate the changes that occur
when a small severing rate is included, we perturb these solutions
as follows
\begin{align}
f^+(x) &= \mathrm{e}^{-(1-r v)x} (1+ s \hat{f}^+(x)) + O(s^2)\\
f^-(x) &= v \mathrm{e}^{-(1-r v)x} (1+ s \hat{f}^-(x))+ O(s^2).
\end{align}
Inserting these expressions into equations \eqref{eq:sever-SSall}
and dropping all higher order terms gives a set of equations that
can be solved to yield
\begin{align}
f^+(x) =& e^{-(1-r v)x} \left(1+ s \left[  \frac{1+r v^2}{(1-r v)^2}x -\frac{1+r v^2}{2(1-r v)}x^2 \right] \right)\nonumber\\& + O(s^2)\\
f^-(x) =& v e^{-(1-r v)x} \biggr(1+ s \biggr[ \frac{1+v}{(1-rv)^2} +
\frac{-v+r v+2 r v^2}{(1-r v)^2}x \nonumber\\
& -\frac{1+r v^2}{2(1-r v)}x^2 \biggr] \biggr)  + O(s^2)
\end{align}
We note that both solutions will become negative for large values of $x$. However, even though this is decidedly unphysical, the
effect on measurable parameters such as the average microtubule
length is small (for small $s$), because of the rapid decay of $|f^+(x)|$ and $|f^-(x)|$. The properties of the resulting distributions, such as the average length, will therefore still remain valid, subject to the bounds on $s$ computed below.

The first-order distributions in $s$ satisfy the total microtubule
number constraint $\integral{[f^+(x)+f^-(x)]}{x}{0}{\infty} =
(1+v)/(1- r v)  + O(s^2)$, consistent with the general result
\eqref{eq:totalnumber}. For the average length we obtain (to first
order in $s$)
\begin{equation}
\bar{x} = \frac{1}{1-r v} - s \frac{1+v}{(1-r v)^4}  +
O(s^2)\mathrm{.}\label{eq:smallk-x}
\end{equation}
For $r=0$, this is consistent with the exact result \eqref{eq:avLengthR0}. The resulting (linear) predictions are indicated in figures \ref{fig:severingR0} and \ref{fig:severingR1}.

Finally, we determine the length variation
\begin{equation}
\sigma_x^2 = <(x-\bar{x})^2>=\frac{1}{(1-r v)^2} - k \frac{2(2+v+r
v^2)}{(1-r v)^5} + O(s^2)
\end{equation}
and, from that, the coefficient of variation ($\sigma_x/\bar{x}$)
\begin{equation}
\sigma_x/\bar{x} = 1 - k \frac{1+r v^2}{(1-r v)^3} + O(s^2).
\label{eq:smallk-sx}
\end{equation}
This number provides a measure for the relative width of the
distribution. The results \eqref{eq:smallk-x} and
\eqref{eq:smallk-sx} confirm that severing decreases both the
weighted average and the relative width of the length distribution.

The results above have been obtained under the assumption that $s$
is very small. To make an \emph{a priori} estimate for the
validity range of $s$, we determine the relative importance of the
terms on the right-hand side of equations \eqref{eq:sseqn1} and
\eqref{eq:sseqn2}. Using the results in the absence of severing as
a benchmark, the terms \emph{not} involving $s$ give contributions
of the order $(1-r v)e^{-(1-r v)x}$. Comparing the integral term
(evaluated for the $s=0$ situation) with this term gives $s \ll
(1-r v)^2 $. The term that is proportional to $s x$ will dominate
the other terms for large $x$, but this does not significantly
affect the results if it only occurs for lengths that are much
longer than the average length. Evaluating the terms at
$x=n/(1-rv)$, where $n$ is the number of average lengths, we
obtain the constraint $s \ll (1-rv)^2/n$. Because $n$ is of the
order 1, we simply state that the approximation is accurate for
$s \ll (1-r v)^2$.

\end{document}